\documentclass[twocolumn,showpacs,preprintnumbers,amsmath,amssymb,final]{revtex4}

\usepackage{graphicx}
\usepackage{dcolumn}
\usepackage{bm}

\def\BEq{\begin{equation}}
\def\EEq{\end{equation}}
\def\BEqA{\begin{eqnarray}}
\def\EEqA{\end{eqnarray}}
\def\BEn{\begin{enumerate}}
\def\EEn{\end{enumerate}}
\def\BWT{\begin{widetext}}
\def\EWT{\end{widetext}}

\def\s{\sigma}

\def\dag{\dagger}
\def\adj{{^{\dag}}}

\begin{document}


\title{Controlled-NOT logic with nonresonant Josephson phase qubits 
}

\author{Andrei Galiautdinov}
 \email{ag@physast.uga.edu}
\affiliation{
Department of Physics and Astronomy,
University of Georgia, Athens, Georgia
30602, USA
}

\date{\today}

\begin{abstract}
We establish theoretical bounds on qubit detuning for high fidelity controlled-NOT 
logic gate implementations with weakly coupled
Josephson phase qubits. It is found that the value of qubit detuning during the 
entangling 
pulses must not 
exceed $2g$ for two-step, and $g$ for single-step control sequences, where 
$g$ is the relevant coupling constant.
\end{abstract}


\maketitle


\section{\label{sec:INTRO}Introduction}

In our previous work
on steering with Josephson phase qubits \cite{MYCNOTPAPER1} 
we found two peculiar
single-step controlled-NOT (CNOT) implementations involving off-resonance qubits 
detuned by the amount smaller than the characteristic coupling constant. Does the 
detuning always have to be so small? What happens if it gets larger?

These questions seem to be important on several counts.

Firstly, the majority of 
entangling gate designs proposed in the literature assume resonant qubits 
\cite{YAMAMOTO2003, BERKLEY2003, CORATO2003, NORI2005, STEFFEN2006, 
MOOIJ2007, HILL2007, MAXENT2008}.
(However, see Refs. \cite{DEVORET2005, CHALAPAT2008} for notable exceptions.) 
This is hardly surprising since in the rotating wave approximation (RWA)
typically used to analyze superconducting qubits the resonant condition leads to 
relatively simple and easily solvable Hamiltonians containing no local $\sigma^z_k$ terms. 
This works well when the system consists of only two qubits. 
However, it is reasonable to expect that when thousands of such qubits are assembled 
into an integrated circuit, maintaining them on resonance will become a difficult task. 
How then can we be sure that an architecture involving detuned qubits is 
able to reliably generate universal gates, such as, for example, a CNOT gate?

Secondly, some qubits may be fabricated with various defects preventing them from 
being tuned to resonance exactly. Would they still be usable? And what if for some 
applications it becomes advantageous (or even necessary) to use qubits with 
intrinsically different level splittings?

Thirdly, we should also keep in mind that detuning is routinely used 
in actual experiments to ``decouple'' the qubits in order to perform local 
(that is, nonentangling) operations 
\cite{YAMAMOTO2003, WELLSTOOD2003, PLOURDE2004, DEVORET2007}.
It would be useful if, after such decoupling is performed, we do not have to worry 
about bringing the qubits back to exact resonance when doing 
subsequent entangling operations. 

Additionally, from the purely theoretical viewpoint,
decoupling provides a useful limit against which to check our calculations.
If at larger detuning the interaction is expected to lose its entangling properties, 
we must be able to predict a crossover into the regime when the gate fidelity 
(of the CNOT, in our case) starts to deteriorate. Thus, the primary goal of this 
paper will be to establish the exact conditions under which such crossover occurs 
for previously proposed CNOT implementations involving superconducting qubits.
   
The paper is organized as follows:

In Section \ref{sec:HAMILTONIAN} we introduce the Hamiltonian
for capacitively and inductively coupled Josephson phase qubits.
Since there is an infinite number of possible local drives and rotating frames to 
choose from, it will be necessary to limit our discussion to situations that are 
simple and yet powerful enough to provide 
some new and interesting physical insights. Our specific choices for Rabi 
pulses and rotating frames will be made in Eqs. (\ref{eq:iH}), (\ref{eq:hatHRWA}), 
and (\ref{eq:HRWA}) of that Section.

In Section \ref{sec:2step} we generalize to finite detuning the familiar two-step 
CNOT sequence involving a local $\pi$-pulse sandwiched between two entanling 
operations. The exact bound on detuning that guarantees generation of the 
perfect (in the RWA) controlled-NOT logic gate will then be given in Eq. (\ref{eq:bound2}).

In Section \ref{sec:1step} we numerically solve the single-step case. We will see 
that in that case the restriction on detuning for perfect CNOT generation is 
somewhat stronger than in the two-step case (Eq. (\ref{eq:bound1})).

In Section \ref{sec:largedetuning} we present optimal results for approximate 
CNOT gates at detuning larger than maximally allowed. We will see how the 
Makhlin invariants and fidelity of the optimized gates deviate from their ideal CNOT values. 
We will also simulate the Weyl chamber steering trajectories corresponding to 
such optimized gates.

We conclude in Section \ref{sec:conclusion} with a brief summary of our results.

\section{Notation}

In what follows, we will use the notation that is convenient for Lie algebraic manipulations,
\BEqA
&& X_k = \frac{i}{2}\s_k^x, \; Y_k =  \frac{i}{2}\s_k^y, \; Z_k =  \frac{i}{2}\s_k^z,
\quad (k = 1,2) ,\nonumber \\
&&  XX =  \frac{i}{2}\s_2^x\s_1^x, 
\; YY =  \frac{i}{2}\s_2^y\s_1^y,
\; ZZ =  \frac{i}{2}\s_2^z\s_1^z, \nonumber \\
&& XY= \frac{i}{2}\s_2^x\s_1^y,  \; YX = \frac{i}{2}\s_2^y\s_1^x.
\EEqA
Notice, that $[XX,YY]=[XY,YX]=0$, and $ZZ$ commutes with each of the 
$XX$, $YY$, $XY$, $YX$ operators. 

Why is this notation convenient?

Consider the following transformation (called ``going to a rotated frame'') on 
the Lie algebra $su(4)$ of the two-qubit system:
\BEq
\label{eq:lierotation}
XX \longrightarrow e^{-\theta Z_1} XX e^{\theta Z_1} = XX \cos \theta  + XY \sin \theta .
\EEq
This transformation can be nicely interpreted as a rotation of vector $XX$ by an 
angle $\theta$ in the {\it real} 
vector space spanned by the generators of the group SU(4). This is how the 
continuous group acts on its Lie
algebra. Mathematicians call it the adjoint representation. The algebra plays 
the role of the 
representation space for its own group.

Since we do work at the level of algebra, and not at the level of the Hilbert space when 
discussing equivalence classes of gates, this is a very convenient notation. It simplifies 
things. Also notice how naturally the periodicity of $2\pi$ appears at this level of description.

Now, if we were to write the same transformation in terms of the Pauli matrices, we would 
have to remember to put in the imaginary unit $i$ and the factors of 1/2 in the exponents
on the left hand side of Eq. (\ref{eq:lierotation}),
\[
e^{-i\theta\sigma_z^1/2} (\sigma_2^x \sigma_1^x) e^{i\theta\sigma_z^1/2}= 
(\sigma_2^x \sigma_1^x)\cos \theta + (\sigma_2^x \sigma_1^y) \sin \theta,
\]
breaking its beautiful symmetry.

\section{The Hamiltonian}
\label{sec:HAMILTONIAN}

When restricted to the computational subspace, the Hamiltonian for 
two coupled Josephson phase qubits, one of which is driven by a resonant rf pulse,
is given by
\BEqA
\label{eq:iH}
iH(t) &=& -\omega Z_1 -(\omega+\delta) Z_2 \nonumber \\
&& +  2 \Omega_1\cos\left(\omega t\right)X_1 +  2(g YY + \tilde{g} ZZ),
\EEqA
where $g, \tilde{g}\ll\omega$ are the coupling constants, $\omega$ is the level 
splitting of the 
first qubit, $|\delta|\ll \omega$ is the detuning, and $\Omega_1$ is the 
corresponding Rabi frequency. 
Here we assume that $g \simeq \Omega_1$, which
differs from the condition $g \ll \Omega_1 \ll \omega$ adopted in Ref. \cite{DEVORET2005}. 
For realistic systems, $\omega \approx 10$ GHz,
$g\approx 10$ MHz. For capacitive coupling, $\tilde{g}=0$; for inductive 
coupling, $\tilde{g}\lesssim 0.1 g$. 

 In the doubly rotating frame defined by 
 $e^{iH_0t}(\dots) e^{-iH_0t}$, with $iH_0 \equiv -\omega (Z_1 + Z_2)$, after 
 averaging over fast oscillations,
the system Hamiltonian is time-independent,
\BEqA
\label{eq:hatHRWA}
iH_{\rm RWA}  & = &   -\delta Z_2 + \Omega_1 X_1 + 
i{\cal H} + \tilde{g}ZZ,
\EEqA
with
\BEqA
i{\cal H}  =  g(XX+YY) =
i g \begin{pmatrix} 
0 & 0 & 0 & 0 \cr 
0 & 0 & 1 & 0 \cr 
0 & 1 & 0& 0 \cr 
0 & 0 & 0 & 0 \end{pmatrix}.
\EEqA

Alternatively, to perform a useful consistency check, we consider another rotating 
frame defined by
$e^{i{\hat H}_0t}(\dots) e^{-i{\hat H}_0t}$, with
$i{\hat H}_0 = -\omega Z_1 -(\omega+\delta) Z_2$. In this frame, the RWA Hamiltonian 
is
\BEqA
\label{eq:HRWA}
i{\hat H}_{\rm RWA}(t)  &=&   \Omega_1 X_1 + i {\hat {\cal H}}(t) + \tilde{g}ZZ ,
\EEqA
where now we have a slowly varying interaction term given by
\BEqA
\label{eq:timedependentinteraction}
i{\hat {\cal H}}(t) &=&  g\left[ (XX+YY) \cos (\delta t) + (YX-XY)\sin (\delta t)\right]
\nonumber \\
&=&
i g \begin{pmatrix} 
0 & 0 & 0 & 0 \cr 
0 & 0 & e^{-i\delta t} & 0 \cr 
0 & e^{+i\delta t} & 0& 0 \cr 
0 & 0 & 0 & 0 \end{pmatrix} .
\EEqA
The central block of this matrix has the form of a rotating drive
for a spin-1/2 system for which the analytical solution is well known
\footnote{
Recall that for a two-level system driven by the Hamiltonian
$H = \omega_0\sigma^z/2  + g (\sigma^x \cos (\delta t) 
+ \sigma^y \sin (\delta t))$, 
the corresponding time evolution operator is given by
$U(t) = e^{-it\delta \sigma^z/2}
e^{-it((\omega_0-\delta)\sigma^z/2 +g \sigma^x)}$. 
Eq. (\ref{eq:timedependentinteraction}) corresponds to
$\omega_0=0$. }. This observation will prove helpful for
calculations in Section \ref{sec:2step}. 

We will now show how these two RWA Hamiltonians lead to locally equivalent
 CNOT implementations.

\section{Two-step CNOT}
\label{sec:2step}

The well-known two-step CNOT implementation for {\it resonant} 
($\delta = 0$) qubits
has control sequence \cite{BURKARD, ZHANG}
\BEqA
\label{eq:2stepCNOT}
{\rm CNOT}_{(2)} = e^{i(\pi/4)} R_{\rm post} 
\left[
U(t_{(2)}) e^{-\pi X_1} U(t_{(2)}) 
\right]
R_{\rm pre},
\EEqA
with $t_{(2)} = \pi/4g$, where 
\BEqA
U(t_{(2)}) &=& e^{- t_{(2)} (g(XX+YY)+\tilde{g}ZZ)}\nonumber \\
&=&e^{- (\pi\tilde{g}/4g)ZZ}
\begin{pmatrix}1 &    0& 0& 0 \cr
 0& 1/\sqrt{2} & -i/\sqrt{2} &       0 \cr
 0&  -i/\sqrt{2} &       1/\sqrt{2} &       0\cr
 0 &0&  0 &1 \end{pmatrix},
\EEqA
and $R_{\rm post, pre}$ are some local rotations. We will choose
\BEq
\label{eq:Rprepost0}
R_{\rm post}=e^{-(\pi/2)Y_2}, \;
R_{\rm pre} = e^{-(\pi/2)Z_2}e^{+(\pi/2)(X_2+X_1)}.
\EEq

Of particular importance to us is the entangling part
$U(t) e^{-\pi X_1} U(t)$ that determines the local equivalence class of 
the full gate.
In our case the local class is controlled-NOT whose canonical representative
in the computational basis is defined to be
\BEq
{\rm CNOT}  \equiv 
 \begin{pmatrix} 
1 & 0 & 0 & 0 \cr 
0 & 1 & 0 & 0 \cr 
0 & 0 & 0 & 1 \cr 
0 & 0 & 1 & 0 \end{pmatrix}\in {\rm U(4)}, \; {\rm det}({\rm CNOT}) = -1.
\EEq

We ask the following question: 
If $|\delta| > 0$ ({\it detuned} qubits), can we still use the sequence in 
Eq. (\ref{eq:2stepCNOT}) 
to generate a CNOT gate, possibly with {\it different}  gate time and 
different pre- and 
post-rotations? The answer to this question turns out to be ``Yes'',
provided $\delta$ is restricted in a certain way. 

Setting $\Omega_{1}=0$ in Eqs. (\ref{eq:hatHRWA}) and 
(\ref{eq:HRWA}) gives
the time evolution operator
\BEqA
U(t) = e^{- t \tilde{g}ZZ}
\begin{pmatrix}e^{i\delta t/2} &    0& 0& 0 \cr
 0& u  & -iv    &       0 \cr
 0&  -iv &   u^{*} &       0\cr
 0 &0&  0 &e^{-i\delta t/2} \end{pmatrix} 
\EEqA
in frame 1, and 
\BEqA
{\hat U}(t) = e^{- t \tilde{g}ZZ}
\begin{pmatrix}1 &    0& 0& 0 \cr
 0& u e^{-i\delta t/2} & -iv e^{-i\delta t/2}   &       0 \cr
 0&  -ive^{i\delta t/2} &   u^{*}e^{i\delta t/2} &       0\cr
 0 &0&  0 &1 \end{pmatrix} 
\EEqA
in frame 2, where
\BEq
u = \cos\left(\frac{\sqrt{\delta^2+4g^2}}{2}t\right)
+\frac{i\delta}{\sqrt{\delta^2+4g^2}}
  \sin\left(\frac{\sqrt{\delta^2+4g^2}}{2}t\right),
\EEq
\BEq
v = \frac{2g}{\sqrt{\delta^2+4g^2}}  
\sin\left(\frac{\sqrt{\delta^2+4g^2}}{2}t\right).
\EEq
In both frames the Makhlin invariants \cite{MAKHLIN} 
of $U(t) e^{-\pi X_1} U(t)$ are 
\BEqA
G_1 &=&
\left(
\frac{
\delta^2+8g^2\cos^2\left(\frac{\sqrt{\delta^2+4g^2}}{2}t\right)
-4g^2}{\delta^2+4g^2}
\right)^2, 
\EEqA
and
\BEqA
G_2 &=&
(
3\delta^4+
8\delta^2g^2 \left[ 1+ 2\cos(t\sqrt{\delta^2+4g^2})\right] 
\nonumber \\
&&+16g^4\left[ 2 + \cos(2t\sqrt{\delta^2+4g^2})\right]
)/\left(\delta^2+4g^2\right)^2,
\nonumber \\
\EEqA
and are independent of the $ZZ$ coupling. This shows that 
for any $t$ the resulting gates are represented by the same point on the
$(XX, YY)$-plane of the Weyl chamber 
(see \cite{MYCNOTPAPER1, ZHANG} for discussion). 
Since CNOT class corresponds to $G_1 = 0$, $G_2=1$, we get
\BEqA
\label{eq:t2}
t_{(2)} =  \frac{\pi-\arccos\left(\delta^2/4g^2\right)}{\sqrt{\delta^2 + 4g^2}},
\EEqA
with the limit $t_{(2)} \rightarrow \pi/4g$ trivially recovered for vanishing detuning.

For example, for $\tilde{g}=0$, $\delta = 1.00g$, we get
$t_{(2)} = 1.0383\pi/4g$. The corresponding CNOT gate is given by Eq. (\ref{eq:2stepCNOT}), where now
\BWT
\BEqA
&& U(t_{(2)}) = 
\begin{pmatrix}
0.9180 + 0.3965i    &    0       &           0        &          0     \cr    
        0        &     0.6124 + 0.3536i & - 0.7071i  &      0       \cr   
        0         &     - 0.7071i  & 0.6124 - 0.3536i    &    0          \cr
        0          &        0        &          0       &      0.9180 - 0.3965i
         \end{pmatrix}, \nonumber \\
&&R_{\rm post}=e^{-(\pi/2)Y_2}e^{-(\pi/2)(\alpha_2 Z_2 + \alpha_1 Z_1)}, \quad
R_{\rm pre} = e^{-(\pi/2)\left((1+\alpha_2)Z_2 + \alpha_1 Z_1\right)}e^{+(\pi/2)(X_2+X_1)},
\EEqA
\EWT
with $\alpha_2 = 0.5929$, $\alpha_1 = 0.2596$ in frame 1, and
\BEqA
&& {\hat U}(t_{(2)}) = 
\begin{pmatrix}
   1               &   0              &    0       &           0\cr          
        0            & 0.7024 + 0.0817i  &-0.2804 - 0.6491i  &      0\cr          
        0            & 0.2804 - 0.6491i  & 0.7024 - 0.0817i &       0     \cr     
        0             &     0  &                0   &          1      
         \end{pmatrix}, \nonumber \\
&&R_{\rm post}=e^{-(\pi/2)Y_2}e^{-(\pi/2)\tilde{\beta} Z_2}, \nonumber \\
&&R_{\rm pre} = e^{-(\pi/2)(1+\beta)Z_2}e^{+(\pi/2)(X_2+X_1)},
\EEqA
with $\tilde{\beta} = -0.1858$, $\beta = 0.3333$ in frame 2.
Here, the pre- and post-rotations (cf. Eq. (\ref{eq:Rprepost0}))
needed to generate the perfect
in the rotating wave approximation canonical controlled-NOT gate
have been found numerically. Notice that the two-step control 
sequence works only for
\BEq
\label{eq:bound2}
|\delta|\leq 2g \ll \omega.
\EEq

\section{Single-step CNOT}
\label{sec:1step}

Here, for simplicity, we limit our discussion to the RWA Hamiltonian given 
in Eq. (\ref{eq:hatHRWA}) with $\tilde{g}=0$ 
(capacitive coupling). [In this case, a single Rabi term
suffices to implement a CNOT gate. When $\tilde{g}\neq0$, an additional 
local Rabi drive 
$\sim \Omega_2 X_2$ must be applied to the second qubit.]
The CNOT sequence \cite{MYCNOTPAPER1, MYCNOTPAPER2} 
is then
\BEqA
\label{eq:1stepCNOT}
{\rm CNOT}_{(1)} = e^{i(5\pi/4)} R_{\rm post} U(t_{(1)}) R_{\rm pre}, 
\EEqA
where
$t_{(1)} = \pi/2g$,
$\Omega_1 = g\sqrt{(4n)^2-1}$, $n = 1,2,3, \dots$,
\BEqA
U (t_{(1)}) &=& e^{- t_{(1)} [\Omega_1 X_1 + g(XX+YY)]}
\nonumber \\
&=& \frac{(-1)^n}{\sqrt{2}}
 \begin{pmatrix}
1 &    0& 0& -i \cr
 0& 1 & -i   &       0 \cr
 0& -i &   1 &       0\cr
 -i &0&  0 &1
 \end{pmatrix},
\EEqA
and
\BEq
\label{eq:Rprepost15}
R_{\rm post}=e^{-(\pi/2)Y_2}, \;
R_{\rm pre} = e^{-(\pi/2)Z_2}e^{+(\pi/2)(X_2-X_1)}.
\EEq
In order to generalize this single-step implementation to finite detuning, 
we 
optimize the gate parameters $t_{(1)}$ and $\Omega_1$ of
\BEqA
\label{eq:optimizedgate1}
U (t_{(1)}) = e^{- t_{(1)} [-\delta Z_2 + \Omega_1 X_1 + g(XX+YY)]}
\EEqA
using Nelder-Mead simplex direct search with bound constraints for the 
minimum of the distance from the CNOT class defined by
\BEqA
\label{eq:makhlinnorm}
d^2(\Omega_1, t_{(1)}) &:=& |G_1(\Omega_1, t_{(1)})-G_1({\rm CNOT})|^2 
\nonumber \\
&& + |G_2(\Omega_1, t_{(1)})-G_2({\rm CNOT})|^2,
\EEqA
with $G_{1,2}(\Omega_1, t_{(1)})$ being the Makhlin invariants of $U(t_{(1)})$. 
It is important to keep in mind that the distance function 
introduced in Eq. (\ref{eq:makhlinnorm})
is {\it not} a measure of the gate fidelity. 
Infinitely many gates --- all differing from each other by arbitrary local rotations 
--- may have the same value of $d^2(\Omega_1, t_{(1)})$. Once the entangling 
part $U (t_{(1)})$ is found to have $d^2(\Omega_1, t_{(1)})=0$, it can then be 
made 
into the canonical CNOT gate by additional local rotations.

Actual experiments motivate this choice of the distance function. It is generally 
believed that doing local rotations is easy, but performing entanglement is 
difficult. Thus, if by using experimentally available interaction and the local 
controls we can somehow 
steer the system into the ``right'' equivalence class, then making the actual target 
gate will be relatively straightforward.

Performing the optimization we find that 
in order to generate the exact CNOT in the single-step case the detuning 
must be restricted by
\BEq
\label{eq:bound1}
|\delta| \leq g \ll \omega.
\EEq

For example, for maximally allowed $\delta = g$, we get 
$t_{(1)} = 1.2753\pi/2g$, 
$\Omega_1 = 3.7781g$. The resulting gate is given by Eq. (\ref{eq:1stepCNOT}) 
with
\BWT
\BEqA
&& U (t_{(1)}) = 
 \begin{pmatrix}
  -0.2553 - 0.4300i  & 0.4821 - 0.1324i & -0.4821 + 0.1324i  & 0.5001i \cr
   0.4821 - 0.1324i  &-0.0001 + 0.5001i & 0.5001i &  0.4821 + 0.1324i \cr
  -0.4821 + 0.1324i &0.5001i & -0.0001 - 0.5001i  & 0.4821 + 0.1324i \cr
   0.5001i  & 0.4821 + 0.1324i &  0.4821 + 0.1324i  &-0.2553 + 0.4300i
 \end{pmatrix}, 
 \nonumber \\
&&R_{\rm post}=e^{-(\pi/2)Y_2}e^{-(\pi/2)(\alpha_2 Z_2 + \alpha_1 Z_1)}, \quad
R_{\rm pre} = e^{-(\pi/2)\left((1+\alpha_2)Z_2 + \alpha_1 Z_1\right)}
e^{+(\pi/2)(X_2-(1+\gamma_1) X_1)},
\EEqA
\EWT
where $\alpha_2 = 0.8294$, $\alpha_1 = -0.1705$, $\gamma_1 = -0.9998$.

\begin{figure}
\includegraphics[angle=0,width=1.00\linewidth]{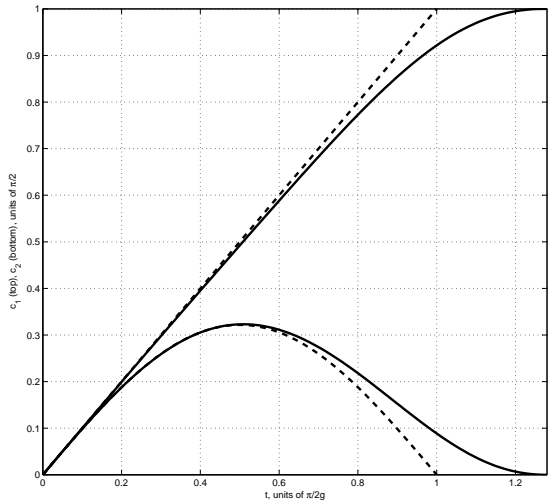}
\caption{
\label{fig:1} 
Time dependence of $c_1$ (top curve) and $c_2$ (bottom curve) for 
single-step 
CNOT implementation 
with capacitively coupled Josephson phase qubits at maximal detuning, 
$\delta = g$. Here, $c_3=0$ at all times.
Dashed curves represent the resonant case, $\delta = 0$.  }
\end{figure}

\begin{figure}
\includegraphics[angle=0,width=1.00\linewidth]{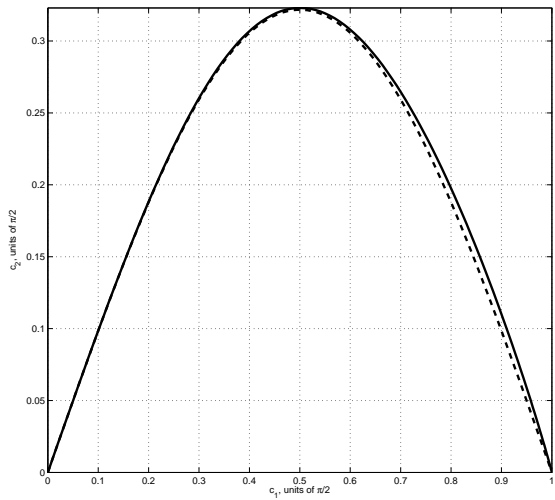}
\caption{
\label{fig:2} 
Single-step steering trajectory corresponding to Fig. \ref{fig:1} that generates
CNOT class at maximal detuning, $\delta = g$.
The steering parameters are given in units of 
$\pi/2$. Dashed curve is for $\delta = 0$.
}
\end{figure}

To visualize how this gate is reached in the course of the unitary evolution 
we simulate the Weyl chamber steering trajectory
$\vec{c}(t) = (c_1(t), c_2(t), c_3(t))$
for the above mentioned values of the gate parameters. 
The goal here is to establish a correspondence \cite{ZHANG},
\BEqA
U(t) \sim e^{- c_1(t)XX-c_2(t)YY-c_3(t)ZZ},
\EEqA
between the physical gate $U (t)$ and the unphysical matrix exponential 
that formally resides in 
the same local equivalence class as $U(t)$. The time-dependent vector 
$\vec{c}(t)$
then represents the dynamically generated class at every moment of 
system's evolution.
Figures \ref{fig:1} and \ref{fig:2} show how the CNOT class that has 
$\vec{c} = (\pi/2, 0, 0)$ is 
generated
in our single-step example with $\delta = g$.

\begingroup
\squeezetable
\begin{table}
\caption{
\label{tab:table1} 
Generation of ideal controlled-NOT logic ($G_1=0$, $G_2=1$) using 
capacitively coupled Josephson phase qubits at finite detuning.
In the two-step case, $|\delta| \leq 2g \ll \omega$, $\Omega_1 =0$, 
and $t_{(2)} = (\pi/4g)T_{(2)}$, as given in Eq. (\ref{eq:t2}).
In the single-step case, $|\delta| \leq g \ll \omega$. The optimal values 
of $\Omega_1 \approx g\sqrt{15}$ and $t_{(1)} = (\pi/2g)T_{(1)}$
have been found numerically using Nelder-Mead simplex direct search
with bound constraints.}
\begin{ruledtabular}
\begin{tabular}{lccc}
$|\delta|/g$ & $T_{(2)}$ & $T_{(1)}$ & $\Omega_1/g$  \\ \hline

   0.00&   1.0000&   1.0000& 3.8730\\
   0.10&   1.0003&   1.0009& 3.8724\\
   0.20&   1.0014&   1.0037& 3.8707\\
   0.30&   1.0031&   1.0085& 3.8679\\
   0.40&   1.0056&   1.0155& 3.8638\\
   0.50&   1.0088&   1.0253& 3.8583\\
   0.60&   1.0128&   1.0386& 3.8513\\
   0.70&   1.0177&   1.0568& 3.8422\\
   0.80&   1.0235&   1.0827& 3.8303\\
   0.90&   1.0303&   1.1245& 3.8132\\
   1.00&   1.0383&   1.2753& 3.7781\\ \hline
   1.10&   1.0476&&\\
   1.20&   1.0585&&\\
   1.30&   1.0713&&\\
   1.40&   1.0863&&\\
   1.50&   1.1042&&\\
   1.60&   1.1261&&\\
   1.70&   1.1536&&\\
   1.80&   1.1901&&\\
   1.90&   1.2445&&\\
   2.00&   1.4142&&\\
 \end{tabular}
\end{ruledtabular}
\end{table}
\endgroup

For other values of qubit detuning, the gate parameters required to 
implement perfect two-step and 
single-step CNOT logic are listed in Table \ref{tab:table1}.

\section{What happens at larger detuning?}

\label{sec:largedetuning}
 
When detuning exceeds the limits set in Eqs. (\ref{eq:bound2})
and (\ref{eq:bound1}), our implementations begin to deviate from 
their perfect controlled-NOT form.
To visualize how that happens, we again search for the gate parameters 
that produce the gates that belong to the local equivalence classes closest to 
CNOT in the sense of the distance function defined in Eq. (\ref{eq:makhlinnorm}).
 The corresponding results for single-step implementation are given in 
 Table \ref{tab:table2}.   
 
\begingroup
\squeezetable
\begin{table}
\caption{
\label{tab:table2} 
Optimized values of the gate parameters for the gates closest to the single-step 
CNOT in the sense of Eq. (\ref{eq:makhlinnorm}) at $ |\delta| \geq g$. 
Here, $t_{(1)} = (\pi/2g)T_{(1)}$. Notice how the Makhlin invariants deviate from 
their ideal CNOT values for larger $\delta$. 
}
\begin{ruledtabular}
\begin{tabular}{lcccc}
$|\delta|/g$ & $T_{(1)}$ & $\Omega_1/g$ & $G_1$ & $G_2$ \\ \hline
   1.00&   1.2753& 3.7781 & 0.0000 & 1.0000\\ 
   1.10&   1.2330& 3.7470 &0.0030 &  0.9994 \\
   1.20&   1.1945& 3.7323& 0.0106 & 0.9978\\
   1.30&   1.1590 & 3.7250&0.0214 &  0.9955\\
   1.40&   1.1262& 3.7203&0.0340& 0.9927\\
   1.50&   1.0961& 3.7152&0.0476&  0.9898\\
   1.60&   1.0686 & 3.7074  &0.0614 &  0.9867\\
   1.70&   1.0438 & 3.6952  & 0.0749 &  0.9837\\
   1.80&   1.0216 & 3.6772& 0.0879 & 0.9808\\
   1.90&    1.0019 &3.6519& 0.1003&  0.9780\\
   2.00&    0.9849 & 3.6179&0.1118& 0.9754\\
 \end{tabular}
\end{ruledtabular}
\end{table}
\endgroup

\begin{figure}
\includegraphics[angle=0,width=1.00\linewidth]{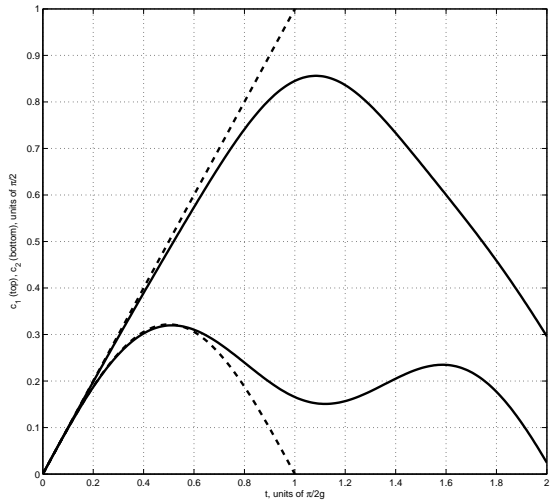}
\caption{
\label{fig:3} 
Time dependence of $c_1$ (top curve) and $c_2$ (bottom curve) for 
single-step implementation generating the class closest to
CNOT in terms of Eq. (\ref{eq:makhlinnorm}) at $\delta = 1.5g$. Here, 
$c_3=0$ at all times.
Dashed curves represent the resonant case.  
Time is measured in units of $\pi/2g$. The closest class is reached at 
$t_{(1)} = 1.0961$.}
\end{figure}

\begin{figure}
\includegraphics[angle=0,width=1.00\linewidth]{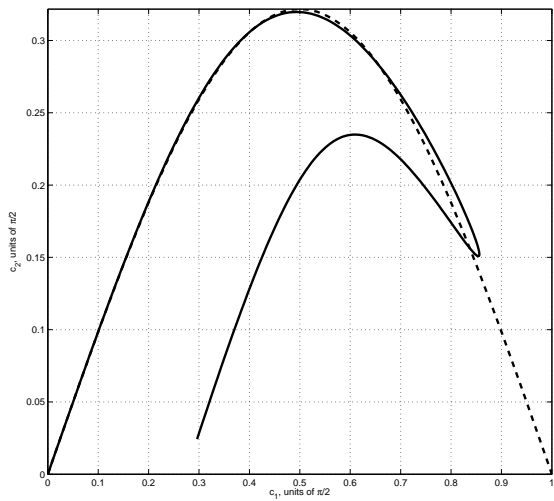}
\caption{
\label{fig:4} 
Single-step steering trajectory corresponding to Fig. \ref{fig:3} at 
detuning $\delta = 1.5g$.
The steering parameters are given in units of 
$\pi/2$. Dashed curve is for $\delta = 0$.
}
\end{figure}
 
 Figures \ref{fig:3} and \ref{fig:4} show the simulated single-step steering 
 trajectory generated by
the RWA Hamiltonian given in Eqs. (\ref{eq:hatHRWA}) and 
(\ref{eq:optimizedgate1})
at $\delta = 1.5g$. 
Direct search for local pulses (not shown) results in the 
optimized gate,
\BEq
U_{\rm opt} = 
 \begin{pmatrix}
0.9866 & - 0.1122i  & 0.0258i  & 0.1158 \cr
-0.1122i  & 0.9866 &  -0.1158 & - 0.0258i \cr
-0.1186 & 0.0009i  & 0.1122i  & 0.9866 \cr
 - 0.0009i  & 0.1186 & 0.9866 & 0.1122i
 \end{pmatrix},
\EEq
whose intrinsic fidelity with respect to the canonical controlled-NOT gate is
\BEqA
{\cal F} &\equiv& \sqrt{1- {\rm tr} \left[ \left(U_{\rm opt} - {\rm CNOT} \right)\adj
\left(U_{\rm opt} - {\rm CNOT} \right)\right]} \nonumber \\
& =& 0.9448.
\EEqA

\section{Conclusion}
\label{sec:conclusion}

In summary, we have demonstrated that the CNOT pulse sequences previously 
proposed for 
resonant Josephson phase qubits may still be used at finite detuning. 
To achieve high fidelity of the resulting gate the value of the detuning during the 
entangling 
operations should not be greater than $2g$ in the two-step implementation and 
$g$ in the single-step implementation, respectively.

\begin{acknowledgments}

This work was supported by IARPA 
under grant W911NF-08-1-0336 and by the NSF under grant CMS- 
0404031. The author thanks Michael Geller and John Martinis
for useful discussions.

\end{acknowledgments}

\end{document}